\documentclass{JHEP3}
\usepackage[dvips]{graphicx}
\usepackage{amsmath,amssymb}

\newcommand{\be}{\begin{equation}}
\newcommand{\ee}{\end{equation}}
\newcommand{\bea}{\begin{eqnarray}}
\newcommand{\eea}{\end{eqnarray}}

\newcommand{\mat}{\left ( \begin{array}{cc}}
\newcommand{\emat}{\end{array} \right )}

\def\Sl#1{\rlap{\raisebox{.15ex}{$\mskip 4 mu /$}}#1}  

\title{Finite-volume Correction to the Pion Decay Constant in the Epsilon-Regime}

\author{Poul H. Damgaard$^a$, Thomas DeGrand$^b$ and Hidenori Fukaya$^a$ \\
\it $^a$ The Niels Bohr Institute and the Niels Bohr International Academy, Blegdamsvej 17, DK-22100 Copenhagen, Denmark \\
        E-mail: \email{phdamg@nbi.dk}, \email{hfukaya@nbi.dk} \\
\it $^b$Department of Physics, University of Colorado, Boulder, CO 80309 USA \\
 Email: \email{degrand@aurinko.colorado.edu}
}

\abstract{In the chiral limit of QCD, the pion decay constant $F$ can be
extracted from lattice gauge theory by means of a coupling to isospin
chemical potential. Here we compute the leading correction due to finite
volume in the $\epsilon$-expansion of chiral perturbation theory. A comparison
is made to recent Monte Carlo data.
}

\preprint{COLO-HEP-532}

\begin{document}

\section{Introduction}

Recently, much effort has gone into numerical lattice gauge theory computations
in QCD very close to the chiral limit. 
Simulations have started
to use chiral fermions for the dynamical quarks of masses
close to the physical values \cite{DeGrand,Fukaya,Hasenfratz}.
It is thus becoming increasingly
important to explore the analytical tools available for extracting physical
observables in this region of almost massless quarks. Here we shall focus on
one of the most important low-energy constants of QCD, the pion decay
constant $F$. Because we consider the chiral limit, and because our purpose 
is to derive expressions useful for a comparison between finite-volume
lattice gauge theory simulations and analytical predictions, we phrase our
analysis in terms of the so-called $\epsilon$-regime of QCD. Roughly
speaking, we will here perform an expansion in $1/L$, where $L$ is
a typical length scale of the given four-volume $V$, $i.e.$, $L \equiv
V^{1/4}$. This is a finite-volume scaling regime of almost-massless QCD.

To extract the pion decay constant $F$ from lattice data one would like
to focus on an observable that is particularly sensitive to the value of $F$.
An example of such an observable has been given in ref. \cite{DHSS}. It is
a two-point spectral correlation function of the Dirac operator when subjected
to an (imaginary) isospin chemical potential $\mu$. The advantage of this particular
spectral function is that it focuses on a null-effect: when the isospin chemical
potential vanishes, the two-point correlation function has a peak of
zero width at coincident points. As soon as the imaginary isospin chemical
potential is turned on, the width becomes finite. Since it depends only
the scaling variable $\mu^2F^2V$,
this gives  a direct way to measure $F$ from  lattice data. The precise form of
the spectral two-point function has been computed to leading order from chiral 
perturbation theory in the $\epsilon$-regime for both the quenched theory and for 
QCD with two nearly massless quarks. More recently, it has been shown how these 
results can also be derived from a
chiral Random Two-Matrix Theory \cite{ADOS}. The Random Matrix Theory approach
has the advantage of being easily generalizable to any number of light
quark flavors $N_f$, and also to the ``partially quenched'' situation
in which the chemical potential is coupled only to valence quarks and not
to the physical $u$ and $d$ quarks of QCD. In addition, it provides all
spectral correlation functions, of arbitrary order. In this way even
individual eigenvalue distributions can be computed, so that one will
be able to extract both the chiral condensate $\Sigma$ and the pion
decay constant $F$ by a fit to the distribution of  just one single
Dirac operator eigenvalue. Any sector of fixed topological charge $\nu$
can be considered. An alternative using ordinary baryon chemical potential
(which leads to a complex Dirac operator spectrum) has also been
considered \cite{AW}.
The equivalence between spectral correlation functions
of arbitrarily high order computed from either the
chiral Random Matrix Theories and the Chiral Lagrangian
to leading order in the $\epsilon$-regime has recently
been shown by Basile and Akemann \cite{BA}.

When comparing with lattice data it is crucial to be able to estimate the
error due to restricting the analysis to leading order in the $\epsilon$-expansion
of chiral perturbation theory. Here we consider the first correction to
the effective field theory at an external isospin chemical potential.
Actually, the calculation is of more general validity, corresponding
to any vector source $v_{\mu} = v_{\mu}^aT^a$ on $SU(N)$ or $U(N)$.
 
In the next section we briefly review the set-up of the $\epsilon$-expansion
in chiral perturbation theory. In section 3 we present the result of
our calculation and make a comparison with preliminary lattice data.
Section 4 contains our conclusions.  

\section{Chiral Perturbation Theory in the $\epsilon$-Regime of QCD}

We are interested in QCD with two light flavors, but the calculation is
easily done for an arbitrary number of light flavors $N_f$. We likewise
consider isospin chemical
in a more general sense: We couple
all light quarks to a quark (baryon) charge operator ${\mathbf B}$. For two light
flavors we have ${\mathbf B} = \mu\sigma^3$, where $\mu$ is isospin chemical
potential and $\sigma^3$ is the third Pauli matrix. In the more general
case we consider $N_f$ to be even, and ${\mathbf B} = \mu\sigma^3\otimes{\mathbf 1}$.
In fact, all that is important for the calculation that follows is
the condition of vanishing trace, ${\rm Tr}{\mathbf B} = 0$. Even if
we generalize to a vector source that is not traceless, the correction to
the zero-mode integral is unchanged, and only the constant part of
the action is modified. The calculation therefore also applies to, for
example, ordinary baryon chemical potential.  

In terms of Dirac operators we are thus dealing with two kinds,
\bea
D_+\psi_+^{(n)} &\equiv & [\Sl{D}(A)+i\mu\gamma_0]\psi_+^{(n)} ~=~
 i\lambda_+^{(n)}\psi_+^{(n)}  \cr
D_-\psi_-^{(n)} &\equiv & [\Sl{D}(A)-i\mu\gamma_0]\psi_-^{(n)} ~=~
 i\lambda_-^{(n)}\psi_-^{(n)}
\label{Ddef}
\eea
and correspondingly two sets of eigenvalues $\lambda_{\pm}^{(n)}$. 
Here $\Sl{D}(A)$ is the ordinary Dirac operator and $\psi_\pm^{(n)}$
denotes the eigenfunctions.
The
method for determining the pion decay constant $F$ from these sets of
eigenvalues has been explained in refs. \cite{DHSS,ADOS}. That analysis
was restricted to the tree-level chiral Lagrangian in the $\epsilon$-regime.
In order to use it to compute $F$ from a lattice simulation, one has to
know the size of the leading correction. That is the subject of this investigation.

We are also interested in a situation where the theory under consideration
has $N_f$ light physical flavors that do {\em not} couple to isospin chemical
potential, while $N_v$ light valence quarks do couple to it. Computationally,
this situation is of much interest since it means that the gauge field 
configurations that need to be used are ordinary ones, without any reference
to isospin chemical potential. Only the smallest eigenvalues corresponding to 
valence quark Dirac operators of the kind (\ref{Ddef}) need to be computed anew.

Because we consider the chiral limit and assume that chiral
symmetry is broken spontaneously, we phrase the analysis in terms of the
effective theory of pseudo-Goldstone bosons, the chiral Lagrangian. In the
$\epsilon$-regime of QCD \cite{GL,Neuberger} one performs an expansion
around the zero momentum modes of the Goldstones, taking into account
the non-zero momentum modes in a perturbative manner. Interestingly,
the topological charge of gauge field configurations $\nu$ then plays
a highly non-trivial role \cite{LS}. In a slightly confusing choice
of terminology this is known as the $\epsilon$-expansion of chiral
perturbation theory.
 
Let us consider the $(N_f+N)$-flavor chiral Lagrangian,
\bea
\mathcal{L}
&=&
\frac{F^2}{4}{\rm Tr}
\left[(\nabla_0 U(x))^\dagger\nabla_0 U(x)+
\sum^3_{i=1}(\partial_i U(x))^\dagger\partial_i U(x) \right]
-\frac{\Sigma}{2}{\rm Tr}\mathcal{M}
(U(x)+U(x)^\dagger),
\nonumber\\
\eea
where 
\bea
\nabla_0 U(x)=\partial_0 U(x)-i[{\mathbf B},U(x)].
\eea
Here we assume that the mass matrix $\mathcal{M}$ is diagonal and
all its elements are taken in the $\epsilon$-regime;
$\mathcal{M}\Sigma V\sim {\cal O}(1)$. The $N$ additional replicated
flavors can be used to obtain the pertinent expression for
the partially quenched theory, after embedding $N_v$ valence
quarks into these $N$ flavors and taking the replica limit
$N \to 0$.  We have singled out the zero-component of
$\nabla_{\mu}$ in order to keep the direct connection to 
isospin chemical potential. As explained in the introduction,
this restriction is actually immaterial, and a more general
traceless vector source will lead to identical results.

Separating the zero-mode from non-zero modes (denoted by $U$ and
$\xi(x)$ respectively),
\bea
U(x)=U\exp(i\sqrt{2}\xi(x)/F),
\eea
the partition function in a sector 
with a fixed topological charge $\nu$ is written
\bea
\mathcal{Z}^\nu_{N_f+N} &=& \int_{U(N_f+N)}
dU  (\det U)^\nu \exp \left[
\frac{\Sigma V}{2}{\rm Tr}[\mathcal{M} U+\mathcal{M}U^\dagger]
+\frac{F^2 V}{4}{\rm Tr}[U,{\mathbf B}][U^\dagger, {\mathbf B}]\right]
\nonumber\\
&&\times\int_{U(N_f+N)}d\xi
\exp\left[-\int d^4x\left(\frac{1}{2}{\rm Tr}[\partial_\mu \xi(x)
\partial_\mu \xi(x)]
+\frac{\Sigma}{2F^2}{\rm Tr}[\mathcal{M}(U+U^\dagger)\xi^2(x)]
\right.
\right.
\nonumber\\
&&
\left.
\left.
+\frac{1}{2}{\rm Tr}[(U^\dagger {\mathbf B}U)(\xi^2(x){\mathbf B}
-2\xi(x){\mathbf B}\xi(x)+{\mathbf B}\xi^2(x))]
\right.
\right.
\nonumber\\
&&
\left.
\left.
+{\rm Tr}\partial_0 \xi [{\mathbf B} + U^\dagger {\mathbf B} U,\xi  ]
\right.
\right.
\nonumber\\
&&
\left.
\left.
+\mathcal{L}_q(\xi)
\right)
\right]. \label{action}
\eea
Note that the integrals are performed over the $U(N_f+N)$ group manifold.
The additional kinetic term of the singlet non-zero modes is
\cite{BG}
\bea
\mathcal{L}_{q}(\xi)=\frac{\alpha}{2N_c}(\partial_\mu {\rm Tr}\xi(x))^2
+\frac{m_0^2}{2N_c}({\rm Tr}\xi(x))^2,
\eea
with additional constants $\alpha$ and $m_0$ that are needed when $N_f=0$. 
Here $N_c$ denotes the number of colors, and $1/N_c$ can be usefully
thought of as an expansion parameter. The Lagrangian exhibited in (\ref{action})
includes all terms of ${\cal O}(\epsilon^6)$ that have non-vanishing
expectation values 
in the $\epsilon$-expansion,  plus a 
kinetic term of ${\cal O}(\epsilon^5)$ whose expectation value vanishes. 
The expectation value of its square
contributes an order $\epsilon^2$ term to $F$.\footnote{This term was omitted in an 
earlier version of this paper. The correct result was first given by Akemann,
 Basile and Lellouch\cite{Akemann:2008vp}.}
The counting rules are as follows: $\xi \sim 1/L \sim \epsilon$, ${\cal M} 
\sim \epsilon^4$, ${\mathbf B} \sim \epsilon^2$.\footnote{As
is well known \cite{D}, the counting is more involved in the fully quenched
theory due to the additional terms in $\mathcal{L}_q(\xi)$.} The new scaling
variable $F^2\mu^2 V$ is thus of order unity in this counting, just like
${\cal M}\Sigma V$.

The tree-level two-point correlation function of the $\xi$ fields 
can be given compactly for any value of $N_f$, $N_v$ and $N$
\cite{DS}. In the $\epsilon$-regime, 
\bea
\langle \xi_{ij}(x)\xi_{kl}(y)\rangle
&=&
\delta_{il}\delta_{jk}\bar{\Delta}(x-y) 
-\delta_{ij}\delta_{kl}\bar{G}(x-y), \label{DeltaG}
\eea
with $\bar{\Delta}(x)$ and $\bar{G}(x)$ defined by
\begin{eqnarray}
\bar{\Delta}(x) &\equiv& \frac{1}{V}\sum_{p\neq 0}
\frac{e^{ipx}}{p^2},\\
\bar{G}(x) &\equiv& \left\{
\begin{array}{cc}
\frac{1}{N_f}\bar{\Delta}(x) & (N_f\neq 0)\\
\frac{1}{V}\sum_{p\neq 0}
\frac{e^{ipx}(m_0^2+\alpha p^2)/N_c}{p^4} &(N_f=0).
\end{array}
\right. ,
\end{eqnarray}
where the indices $i,j\cdots$ can be taken both in the valence and
sea sectors. 

Quenching artifacts appear in $\bar{G}(x)$
as double poles. One keeps track of the replica limit
$N \to 0$ in the external indices of eq. (\ref{DeltaG}). 
Recently, the chiral expansion
with $\mu \neq 0$ has been considered in the $p$-regime by
Splittorff and Verbaarschot \cite{SV1}. In that case even the propagator
matrix is $\mu$-dependent. Here, to ${\cal O}(\epsilon^4)$ in
the $\epsilon$-expansion these $\mu$-dependent terms do not
contribute. In effect, the propagator we use to compute
the one-loop correction below is insensitive to $\mu$.

\section{The Leading Finite-Volume Correction to $F$}

With the set-up of the previous section it is now straightforward to compute
the leading (one-loop) correction to $F$ in the $\epsilon$-regime. To this
end, we compute the one-loop contribution to the partition function itself:
We bring down the action $S$ and
saturate it to first non-trivial order in the fluctuation field
$\xi(x)$. In this way we obtain the leading finite-volume corrections in the
$\epsilon$-regime. For the term involving the chiral condensate $\Sigma$ this
was done for the full theory in ref. \cite{GL}, and for the quenched and
partially quenched theories in refs. \cite{D,DF}. Here we concentrate
on the term
${\rm Tr}([{\mathbf B},U][{\mathbf B},U^{\dagger}])$.

The computation is now simple. Working in the replica formalism we can in one
sweep compute the correction to $F$ in the full theory, the quenched and
the partially quenched theories\footnote{Here the term ``partially quenched''
is used in the sense described above: We consider $N_v$ valence quarks
coupled to isospin chemical potential, while the $N_f$ physical quarks
do not couple to it.}. We find
\bea
F ~\to~ F\left(1 - \frac{N_f}{F^2} ( \bar{\Delta}(0) -
\frac{1}{V}\int d^4x(\partial_0\bar \Delta(x))^2   )  \right).
\label{Fshift}
\eea
to this order. 
This calculation 
shows
that ${\cal O}(\epsilon^2)$ effects 
can be absorbed in the redefinition of $F$ in the LO Lagrangian,
or the two-matrix theory, analogous to the case of the 1-loop correction to
$\Sigma$ \cite{GL,D}.
Another important observation is that the double-pole terms have canceled.
There is simply no one-loop correction to $F$ in the quenched limit, which
can be viewed as $N_f \to 0$ above. In the partially quenched theory with
both valence and sea quarks in the $\epsilon$-regime the shift in $F$ is
as given in eq. (\ref{Fshift}); there is no difference with the full theory. 
An immediate consequence of this is that
there is to this order no dependence on the poorly determined parameters $m_0$ and
$\alpha$ in the quenched theory. This situation is reminiscent of what
happens with current correlators in the $\epsilon$-regime \cite{DHJLL}.
We can intuitively understand the equality between the full theory and
the partially quenched theory (in the sense defined above) by noting that
the pion loop responsible for the one-loop correction to this order
is insensitive to the value of isospin chemical potential 
$\mu$. It should therefore not matter whether
the gauge field configurations have been generated with dynamical fermions
associated with chemical potential or not.

As is well known, the integrated massless pion propagator $\bar{\Delta}(0)$ is
ultraviolet divergent. In dimensional regularization it actually becomes finite
in four dimensions, and it has been computed for various finite-volume
geometries in ref. \cite{HL}. This can be parametrized in terms of the
so-called ``shape coefficients'' $\beta_1$ and $k_{00}$ which are functions of the
four-geometry only, and which are readily computed for any shape following
the discussion in \cite{HL}. The result is
\bea
F ~\to~ F_L = F\left(1 + \frac{N_f}{F^2}(\frac{\beta_1}{2\sqrt{V}} + \frac{T^2}{2V}k_{00}) \right)
\label{Fshift2}
\eea
for the case $V=L^3T$. The appropriate 
finite-volume propagator has recently been computed by Splittorff and Verbaarschot
in the $p$-regime \cite{SV1} where $\mu$ is of same order as $p$. With
the help of that expression one can find the matching between the
$\epsilon$ and $p$ regimes explicitly.

It is of interest to see the effect of this one-loop shift in $F$ on presently
available data \cite{DeGrand1}.
The one calculation which has been performed to date took $N_f=2$
and used a $12^4$ volume at a lattice spacing where
$a F\sim 0.07$. Using $\beta_1=0.1405$, the correction factor $F_L/F=1.3$.
This is uncomfortably large.
The correction to the condensate is also large,
\bea
\Sigma ~\to~ \Sigma_L = \Sigma\left(1 + \frac{N_f^2-1}{N_f}\frac{\beta_1}{F^2L^2}\right).
\label{Sshift2}
\eea

Because of the weak dependence on simulation volume, it is difficult
to deal with the finite-volume corrections by simply pushing to large volume.
(The simulations of Ref. \cite{Fukaya}, the largest-scale $N_f=2$ simulations done with
fully chiral lattice fermions done to date,
are only done at volumes which are about twice the size of
those of \cite{DeGrand1}.) Precision tests clearly require performing
simulations at several volumes and observing behavior consistent with
Eqs. (\ref{Fshift2}) and (\ref{Sshift2}). Making the simulation volume asymmetric
can help: for example, $\beta_1$ is reduced from 0.1405 to 0.0836 for a lattice of size
$L^3 \times 2L$, as used by Ref. \cite{Fukaya} 
while $k_{00}$ is not changed very much (from 0.070 to 0.083).
The effects of higher-order terms in the chiral expansion will also come in, as $1/V$
corrections \cite{Hansen}. Neither of the volumes of Refs. \cite{DeGrand1} or \cite{Fukaya}
are particularly large by today's standards, but because of their extreme sensitivity
to topology, these kind of $\epsilon$-regime calculations require the use of lattice fermions
with very good chiral properties.
Of course, this problem becomes more acute for larger $N_f$. Here we have considered
QCD with only light $u$ and $d$ quarks.

Finally, we should point out that the present calculation does not directly prove
that the method proposed in ref. \cite{DHSS} can be applied at one-loop level
by just performing the shifts (\ref{Fshift2}) and (\ref{Sshift2}). In principle, 
one would have to perform a full calculation of the needed susceptibility
and resolvent in the graded or replicated chiral Lagrangian,
and then go to the cut at imaginary mass \cite{OSV}. We know of
no indication that the result of such a calculation to one-loop order in
the $\epsilon$-regime should yield a result different from just performing
the shifts (\ref{Fshift2}) and (\ref{Sshift2}), but to our knowledge it has
not been shown explicitly. However, there is much circumstantial evidence
in favor of this being the case. For instance, spectral sum rules \cite{LS} evaluated
to one-loop accuracy are consistent with simply performing such a one-loop
shift. This holds in the theory with finite chemical potential $\mu$ as
well \cite{Luz}.

\section{Conclusions}

We have computed the one-loop correction to $F$ in the $\epsilon$-regime
of QCD with a vector-like source that couples like (imaginary) isospin
chemical potential. This correction is needed in order to quantify the
finite-volume correction to the leading-order formula for $F$ in the $\epsilon$-regime
based on the method described in Ref. \cite{DHSS}. 
Lattice simulations which aim to predict low-energy constants from simulations in the 
$\epsilon-$regime will clearly 
need to be carefully designed to deal with these finite volume corrections.

\vspace{0.5cm}
\noindent
{\bf Acknowledgments:\\}
This work was supported in part by the US Department of Energy (T.D.),
EU network ENRAGE MRTN-CT-2004-005616 (P.H.D.), and a Nishina Fellowship
(H.F.). We are grateful to Kim Splittorff for comments on a preliminary
version of this manuscript. We are also grateful to G.~Akemann, F.~Basile, and L.~Lellouch for
correspondence about an error we made in an earlier version of this paper.

\end{document}